\documentclass[aps,prl,twocolumn,floats,floatfix]{revtex4}

\usepackage{amsmath}
\usepackage{graphicx}
\usepackage{natbib}
\usepackage{latexsym}

\usepackage{ifthen}
\usepackage{ifpdf}
\ifpdf
\usepackage{graphicx}
\usepackage{epstopdf}
\else
\usepackage{graphicx}
\usepackage{epsfig}
\fi

\newcommand{\hpsi}{\hat{\psi}}

\newcommand{\hn}{\hat{n}}

\newcommand{\Lx}{\hat{L}_x}
\newcommand{\Ly}{\hat{L}_y}
\newcommand{\Lz}{\hat{L}_z}

\newcommand{\Lp}{\hat{L}_+}
\newcommand{\Lm}{\hat{L}_-}

\newcommand{\Kx}{\hat{K}_x}
\newcommand{\Ky}{\hat{K}_y}
\newcommand{\Kz}{\hat{K}_z}

\newcommand{\Kp}{\hat{K}_+}
\newcommand{\Km}{\hat{K}_-}
\newcommand{\Kpm}{\hat{K}_\pm}

\newcommand{\hide}[1]{}

\begin{document}

\title{Rapid phase-diffusion between atomic and molecular Bose-Einstein condensates}

\author{I. Tikhonenkov and A. Vardi}
\affiliation{Department of Chemistry, Ben-Gurion University of the Negev, P.O.B. 653, Beer-Sheva 84105, Israel}

\begin{abstract}
We study the collisional loss of atom-molecule coherence after coherently dissociating a small fraction of a molecular Bose-Einstein  condensate into atoms. The obtained $n$-atoms states are two-atom (SU(1,1)) coherent states with number variance $\Delta n\propto n$ compared to $\Delta n\propto \sqrt{n}$ for the spin (SU(2))  coherent states formed by coherent splitting of an atomic condensate.  Consequently, the Lorentzian atom-molecule phase-diffusion is faster than the Gaussian phase-diffusion between separated atomic condensates, by a $\sqrt{n}$ factor.
\end{abstract}

\maketitle

%%%%%%%%%%%%%%%% INTRODUCTION %%%%%%%%%%%%%%%%%%%%%%%%%%

Atom-molecule coherence in a Bose-Einstein condensate (BEC) was first demonstrated experimentally by observing coherent oscillations in a Ramsey-like interferometer \cite{AtomMoleculeRamsey}. Its existence paves the way to a wealth of novel phenomena, including large-amplitude atom-molecule Rabi oscillations \cite{AtomMoleculeRabi}, Atom-Molecule dark states \cite{AtomMoleculeDark}, and 'super-chemistry' \cite{SuperChemistry} characterized by collective, Bose-enhanced and ultraselective dynamics.

One important implication of atom-molecule coherence, is the stimulated dissociation of a molecular BEC into its constituent boson atoms \cite{StimulatedDissociation}. This coherent process is the matter-wave equivalent of parametric downconversion. Like its quantum-optics counterpart, when started from the atomic vacuum (molecular BEC) it involves the hyperbolic amplification of the atom-pair number $n=\langle\hn\rangle$ and of its variance $\Delta n=\left( \langle\hn^2\rangle-\langle\hn\rangle^2\right)^{1/2}$, where $\hn$ is the atomic number operator.  

The exponential growth of $\Delta n$ indicates the formation of a well defined relative-phase $\varphi$ between the molecular BEC and the emerging atomic condensate, as the conjugate phase variance $\Delta\varphi$ is exponentially decreasing. Also like optical parametric amplification, stimulated dissociation is {\it phase-sensitive} for atomic states different than the vacuum state. Given a non-vanishing value of $n$ The relative-phase $\varphi$ between molecules and atoms, determines whether it will be amplified or attenuated.

In this work we propose to use the phase-sensitivity of the stimulated dissociation of a molecular BEC, to implement a sub-shot-noise SU(1,1) interferometer \cite{SU11Interferometer}. The scheme involves two pulses of atom-molecule coupling, separated by a phase-aquisition period, similar to the Ramsey procedure in \cite{AtomMoleculeRamsey} but starting from a {\it molecular} BEC instead of an atomic one. In the limit where the dissociation does not deplete the molecular BEC, the atomic state will be an SU(1,1) or 'two-atom' coherent state (TACS). Our main result is that the $\Delta n\propto n$ atom-number variance of the TACS  results in the loss of atom-molecule phase coherence on a short $\tau_{pd}\propto 1/n$ timescale due to collisional phase-diffusion. By contrast, two initially coherent, separated atomic condensates phase-diffuse on a longer $\tau_{pd}\propto 1/\sqrt{n}$ timescale \cite{PhaseDiffusion}, since their initial state is an SU(2) or 'spin' coherent state (SCS) with $\Delta n\propto \sqrt{n}$. Moreover, we find that for $n\gg 1$ the phase-diffusion of the TACS is Lorentzian in time, as compared to the familiar Gaussian phase-diffusion of the SCS, due to the difference in atom-number distributions between the two coherent states.

%%%%%%%%%%%%%%%%%%%%% HAMILTONIAN %%%%%%%%%%%%%%%%%%%%%%

We consider the atom-molecule model Hamiltonian, where interacting atoms and molecules are coupled by means of either a Feshbach resonance or a resonant Raman transition,
\begin{eqnarray}
\label{HamO}
H&=&E_m\hn_m+E_a\hn+\left(g_{am}\hpsi_m^\dag\hpsi_a\hpsi_a+H.c.\right)\\
~&~&+\frac{u_m}{2}\hpsi_m^\dag\hpsi_m^\dag\hpsi_m\hpsi_m+\frac{u_a}{2}\hpsi_a^\dag\hpsi_a^\dag\hpsi_a\hpsi_a+u_{am}\hn_m\hn,\nonumber
\end{eqnarray}
where $\hpsi_{a,m}$ are the boson annihilation operators for atoms and molecules, $\hn=\hpsi_a^\dag\hpsi_a$, $\hn_m=\hpsi_m^\dag\hpsi_m$ are the corresponding particle numbers, and $E_{a,m}$ are the respective mode energies. The atom-molecule coupling is $g_{am}=|g_{am}|e^{i\phi}$  whereas $u_m,u_a$, and $u_{am}=u_{ma}$ are the collisional interaction strengths for molecule-molecule, atom-atom, and atom-molecule scattering, respectively.

In what follows we shall assume that the molecular condensate remains large and is never significantly depleted by the conversion of a small number of molecules into atoms. This approximation is equivalent to the undepleted pump approximation in parametric downconversion. The molecular field operators $\hpsi_m,\hpsi_m^\dag$ are replaced by the $c$-numbers $\sqrt{n_m}e^{\pm i\phi_m}$ and Eq. (\ref{HamO}) becomes, 
\begin{equation}
H=\delta\Kz+g\Kx+u\Kz^2,
\end{equation}
where $c$-number terms are omitted. Here $\delta=(E_m - 2E_a+2u_{am} n_m-2u_a)$, $g=4|g_{am}|\sqrt{n_m}$, and $u=2u_a$. The operators $\Kp=(e^{i(\phi_m-\phi)}/2)\psi^\dag\psi^\dag$, $\Km=(e^{-i(\phi_m-\phi)}/2)\psi\psi$, $\Kz=\psi^\dag\psi/2+1/4$ are the generators of an SU(1,1) Lie algebra with canonical commutation relations $[\Kp,\Km]=-2\Kz$, $[\Kz,\Kpm]=\pm\Kpm$ and we define the usual Hermitian operators $\Kx=(\Kp+\Km)/2$, $\Ky=(\Kp+\Km)/2i$. Since the Casimir operator of SU(1,1) is ${\hat C}=\Kz^2-\Kx^2-\Ky^2$, we will use for representation the joint eigenstates of ${\hat C}$ and $\Kz$,
\begin{equation}
|k,m\rangle=\sqrt{\frac{\Gamma(2k)}{m!\Gamma(2k+m)}}(\Kp)^m|k,0\rangle
\end{equation}
so that ${\hat C}|k,m\rangle=k(k-1)|k,m\rangle$ and $\Kz|k,m\rangle=(k+m)|k,m\rangle$, with the Bargmann index $k=1/4$ and nonnegative integer $m$. The states $|k,m\rangle$ are atom number states with $n=2m$.

%%%%%%%%%%%%%%%%%%%%%%%
\begin{figure}[t]
\centering
\includegraphics[width=0.4\textwidth]{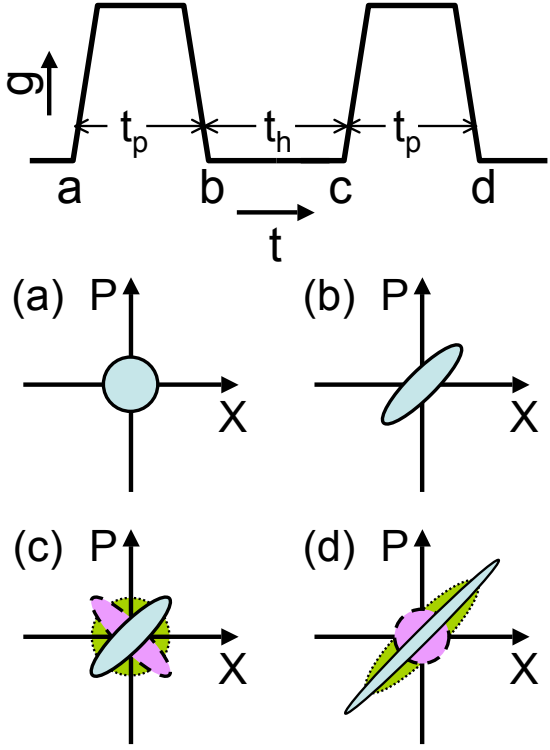}
\caption{(Color online) Atom-molecule SU(1,1) interferometer. The quadrature phase-amplitude distribution is shown at the time points marked on the upper $g(t)$ plot. Note that the polar angle in the $X,P$ plot is $\varphi/2$, not $\varphi$. Starting from the atomic vacuum (a) the first Lorentzian boost results in the squeezing of the atom-molecule phase around $\varphi=\pi/2$ (b), which is allowed to evolve during the hold time (c).  The atom number and its variance after the second pulse (d) depend on the value of $\varphi$ acquired during the hold time. When $\varphi$ remains $\pi/2$ (solid line) the second pulse yields further squeezing with exponentially increasing $n$ whereas if $\varphi=-\pi/2$ (dashed line) the atomic vacuum is recovered. Dotted circles correspond to the loss of coherence due to $\varphi$ phase-diffusion.}
\label{ampd_fig1}
\end{figure}
%%%%%%%%%%%%%%%%%%%%%%

%%%%%%%%%%%%%%%%% INTERFEROMETER %%%%%%%%%%%%%%%%%%%%%%
The SU(1,1) interferometer \cite{SU11Interferometer} for probing the atom-molecule phase coherence, is illustrated in Fig.~\ref{ampd_fig1} by snapshots of the quadrature plane ${\hat X}=\psi+\psi^\dag$, ${\hat P}=(\psi-\psi^\dag)/i$. Starting from the coherent atomic vacuum state $|k,0\rangle$ (Fig.~\ref{ampd_fig1}(a)), the first step is the dissociation of a small fraction of the molecular BEC into atoms, by setting $g\gg \delta, un$. This is attained for Feshbach-coupling, by magnetic control of the atom-molecule detuning and for the optical resonant Raman coupling, by switching the photodissociation lasers. The atomic state following this Lorentzian boost of duration $t_p$, is an SU(1,1) TACS \cite{SU11Interferometer,SU},
\begin{eqnarray}
\label{tacs}
\left|\theta,\varphi\right\rangle_{s}&=&\exp(z\Kp-z^*\Km)|k,0\rangle\\
~&=&\left[1-\zeta^2\right]^{k}\sum_{m}\left[\zeta e^{-i\varphi}\right]^{m}\sqrt{\frac{\Gamma(2k+m)}{m!\Gamma(2k)}}|k,m\rangle~,\nonumber
\end{eqnarray}
with $z=e^{-i\varphi}\theta/2$ and $\zeta=\tanh(\theta/2)$. The obtained squeeze parameter is $\theta=\theta_p\equiv gt$, and the atom-molecule relative phase is $\varphi=\phi-\phi_m+2\phi_a=\pi/2$ (corresponding to quadrature phase of $\pi/4$, see Fig.~\ref{ampd_fig1}(b)). The average atom number of  $|\theta,\varphi\rangle$ is $n=2k(\cosh\theta-1)$ and its variance is $\Delta n=\sqrt{2k}\sinh\theta$ \cite{SU}, corresponding to the amplification of vacuum fluctuations in stimulated dissociation \cite{StimulatedDissociation}.

Next, the coupling $g$ is turned off and the atom-molecule phase is allowed to evolve for a hold-time $t_h$. In the limit where atom-atom and atom-molecule collisions may be neglected ($u=0$), coherence is maintained and the state at the end of the hold time is $\exp(-i\delta\Kz t_h) |\theta_p,\pi/2\rangle=|\theta_p,\pi/2+\varphi_h\rangle$ with $\varphi_h\equiv\delta t_h$ (Fig.~\ref{ampd_fig1}(c)).  The accumulated atom-molecule phase $\varphi_h$ may be determined by a second strong coupling pulse of duration $t_p$ (Fig.~\ref{ampd_fig1}(d)) because the fraction  of reassociated atoms is phase-sensitive \cite{SU11Interferometer}. For example, if $\varphi_h=0$ the second pulse will further dissociate the molecular BEC, whereas if $\varphi_h=\pi$ it will reassociate all atoms into it. The final number of atoms is obtained by noting that the combined boost-rotation-boost sequence $e^{-i\theta_p\Kx}e^{-i\varphi_h\Kz}e^{-i\theta_p\Kx}$ preserves coherence and transforms the vacuum into the final TACS $|\theta_f,\varphi_f\rangle$ with $\cosh\theta_f=[1+\cos\varphi_h]\cosh^2\theta_p-\cos(\varphi_h)$. Hence in the absence of collisions, 
\begin{eqnarray}
\label{nfinal}
n_f&=&2k(\cosh\theta_f-1)=\frac{1+\cos\varphi_h}{2}\sinh^2\theta_p~,\nonumber\\
(\Delta n_f)^2&=&2k\sinh^2\theta_f \\
~&=&\frac{\sinh^2\theta_p}{2}\left[\sin^2\varphi_h+\left(1+\cos\varphi_h\right)^2\cosh^2\theta_p\right]~.
\nonumber
\end{eqnarray}
Note these expressions are slightly different than in Ref. \cite{SU11Interferometer} because the proposed scheme uses two identical, equal phase pulses, as opposed to the reversed Lorentzian boosts of the two degenerate parametric amplifiers in \cite{SU11Interferometer}.  

From Eqs. (\ref{nfinal}) it is clear that an accumulated phase $\varphi_h=\pi$ may be determined within $(\Delta\varphi_h)^2=\left[(\Delta n_f)^2/|\partial n_f/\partial\varphi_h|^2\right]_{\varphi_h=\pi}=(2\sinh^2\theta_p)^{-1}=[8n(n+1)]^{-1}$ accuracy. Thus due to the squeezing inherent in coherent dissociation, $\Delta\varphi_h$ around $\varphi_h=\pi$ goes below the $1/\sqrt{n}$ standard quantum limit (a.k.a. shot-noise limit) and approaches the Heisenberg  $1/n$ uncertainty, where $n$ is the number of atoms dissociated by the first pulse \cite{SU11Interferometer}.
%%%%%%%%%%%%%%%%%%%%%%%%%%%%%%%%%%%%%%%%%%%%%%%%%%%

%%%%%%%%%%%%%%%%%%%%%%%
\begin{figure}[t]
\centering
\includegraphics[width=0.48\textwidth]{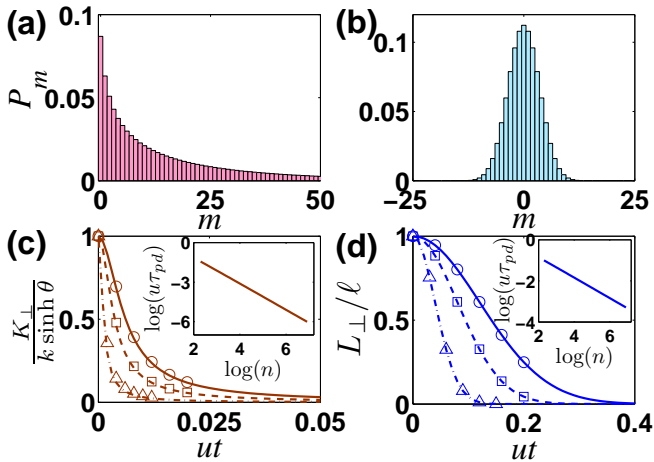}
\caption{(Color online) Comparison of atom-molecule phase-diffusion with the collisional dephasing of separated atomic condensates: (a) number distribution of a TACS $|\theta,\varphi\rangle$ with $\theta=4.8$, corresponding to $n=30$ dissociated atoms; (b) same for a SCS $|\pi/2,\varphi\rangle_s$ with $\ell=n/2=25$; (c) phase-diffusion of TACS with $n=100$ (solid, $\circ$),  167 (dashed, $\Box$), and 500 (dash-dotted, $\triangle$), symbols mark numerical results with $n+2n_m=5000$; (d) same for SCS with $n=70$ (solid, $\circ$), 156 (dashed, $\Box$), and 626 (dash-dotted, $\triangle$), symbols mark numerical results. Insets in (b) and (c) show the decay half-times $\tau_{pd}\propto(un)^{-1}$ for TACS and  $\tau_{pd}\propto (u\sqrt{n})^{-1}$ for SCS.}
\label{ampd_fig2}
\end{figure}
%%%%%%%%%%%%%%%%%%%%%%

%%%%%%%%%%%%%%%%%  PHASE DIFFUSION  %%%%%%%%%%%%%%%%%%%%%%%
Our goal here is to study the effect of interactions on this scenario.  Atom-atom and molecule-atom collisions will degrade atom-molecule coherence during the phase acquisition time since for non-vanishing $u$ the pertinent $|k,m\rangle$  eigenstates are not equally spaced. This collisional dephasing drives the quadrature variances to $(\Delta X)^2=(\Delta P)^2=2n+1$, while keeping $(\Delta X)^2+(\Delta P)^2=2(2n+1)$ fixed, as depicted by the dotted circle in Fig.~\ref{ampd_fig1}(c). Phase information is lost and the final atom number on invoking the second pulse is $\varphi_h$-independent (dotted ellipse in Fig.~\ref{ampd_fig1}(d)). 

Atom-molecule coherence may be quantified by defining the SU(1,1) purity $K^2\equiv\langle \Kz\rangle^2-\langle\Kx\rangle^2-\langle\Ky\rangle^2$. For an SU(1,1) coherent state we have $K=k$ whereas dephasing is characterized by going inside the upper sheet of the hyperboloid $K^2=k^2$, so that $K>k$. Thus, during the $t_h$ hold time where $g=0$ and hence $\langle \Kz\rangle$ is fixed, we may use $K_\perp^2\equiv\langle \Kx\rangle^2+\langle\Ky\rangle^2$ as a measure of coherence. The time dependence of $K_\perp$ is related to the Fourier transform of the initial number distribution. Starting from the TACS $|\theta,\varphi\rangle$ with the number distribution $P_m=|\langle k,m|\theta,\varphi\rangle|^2$ shown in Fig.~\ref{ampd_fig2}(a), we find the exact result that  in the presence of interactions, $K_\perp$ is independent of $\varphi$ , $\delta$ and decays as,
\begin{equation}
K_\perp(t)=\frac{k\sinh\theta}{\left[1+\sin^2 (ut)\sinh^2\theta\right]^{k+1/2}}~.
\label{pdsu11}
\end{equation}
Noting that $\sinh^2\theta=(n/2k)[(n/2k)+2]=4n(n+1)$ we obtain that for a moderately large $n\gg 1$, coherence decays on a $\sin(ut)\sim 1/(2n)$ timescale. Thus we replace $\sinh\theta\approx 2n$, $\sin(ut)\approx ut$ to obtain Lorentzian dephasing $K_\perp=(n/2)[1+(2nut)^2]^{-3/4}$ which reflects the exponential form of $P_m$ and agrees well with numerical simulations (Fig.~\ref{ampd_fig2}(c)). The phase-diffusion time $\tau_{pd}=1/(2un)$ reciprocates the super-Poissonian $\Delta n\propto n$ variance of the TACS.
%%%%%%%%%%%%%%%%%%%%%%%%%%%%%%%%%%%%%%%%%%%%%%%%%%%

%%%%%%%%%%%%%%  COMPARISON TO SU(2)  %%%%%%%%%%%%%%%%%%%%%%%
It is instructive to compare atom-molecule collisional dephasing with phase diffusion between two initially coherent atomic BECs \cite{PhaseDiffusion,PDexperiments,Jo07}. The pertinent Hamiltolian is the two-site Bose-Hubbard model (sometimes referred to as the Bosonic Josephson junction) \cite{BJM} and the initial coherent states are the SU(2) SCS \cite{SU},
\begin{eqnarray}  
\label{coherento}
\left|\theta,\varphi\right\rangle_{s}&=&\exp(z\Lp-z^*\Lm)|\ell,-\ell\rangle\\
~&=&\left[1+\xi^2\right]^{-\ell}\sum_{m=-\ell}^{\ell}\left(\xi e^{-i\varphi}\right)^{\ell+m}
\left(
\begin{array}{c}
2\ell \\ \ell+m
\end{array}\right)^{1/2}|\ell,m\rangle,\nonumber
\end{eqnarray}
where $\xi=\tan(\theta/2)$. The SU(2) generators $\Lx=(\hpsi^\dag_1 \hpsi_2+\hpsi^\dag_2\hpsi_1)/2$,  $\Ly=(\hpsi^\dag_1\hpsi_2-\hpsi^\dag_2\hpsi_1)/(2i)$, and $\Lz=(\hn_1 - \hn_2)/2$, are defined in terms of the boson annihilation and creation operators $\hpsi_i$, $\hpsi^\dag_i$ for particles in condensate~${i=1,2}$ with the number operators $\hn_{i}=\hpsi^\dag_{i}\hpsi_{i}$.  The total particle number $\hn=\hn_1+\hn_2=2\ell$ is conserved and the Fock states $|\ell,m\rangle$ are the standard  ${\hat {\bf L}}^2$, $\Lz$ eigenstates. Experimentally, such states are prepared either by coherently splitting an atomic BEC or by controlling optical or magnetic double-well potentials confining it \cite{PDexperiments,Jo07}. Most common are states with equal population of the two condensates, i.e. $\theta=\pi/2$.

The binomial/Poissonian number distribution of the SCS $|\theta,\varphi\rangle_{s}$ (Fig.~\ref{ampd_fig2}(b)) results in the loss of relative-phase coherence $(L_\perp)^2\equiv\langle\Lx\rangle^2+\langle\Ly\rangle^2$ under a collisional $\delta\Lz+u\Lz^2$ Hamiltonian, as,
\begin{equation}
L_\perp(t)=\ell\sin\theta\left(1-\sin^2(ut)\sin^2\theta\right)^{\ell-1/2}~,
\label{pdsu2}
\end{equation}
approaching for $n\gg 1$, the Gaussian decay $L_\perp=(n/2)\sin\theta e^{-n(\sin\theta ut)^2/2}$ with phase-diffusion time $\tau_{pd}=(u\sin\theta \sqrt{n/2})^{-1}$ \cite{PhaseDiffusion} (Fig.~\ref{ampd_fig2}(d)). For equal $n$, the loss of atom-molecule coherence is thus  typically $\sqrt{n}$ times faster than the phase-diffusion between atomic BECs. We note that the accelerated decay of the super-Poissonian, phase-squeezed SU(1,1) coherent state, is the counterpart of the decelerated phase-diffusion of a sub-Poissonian SU(2) {\it number}-squeezed states, observed experimentally in Ref. \cite{Jo07}.
%%%%%%%%%%%%%%%%%%%%%%%%%%%%%%%%%%%%%%%%%%%%%%%%%%%

%%%%%%%%%%%%%%%%%%%%%%%
\begin{figure}[t]
\centering
\includegraphics[width=0.5\textwidth]{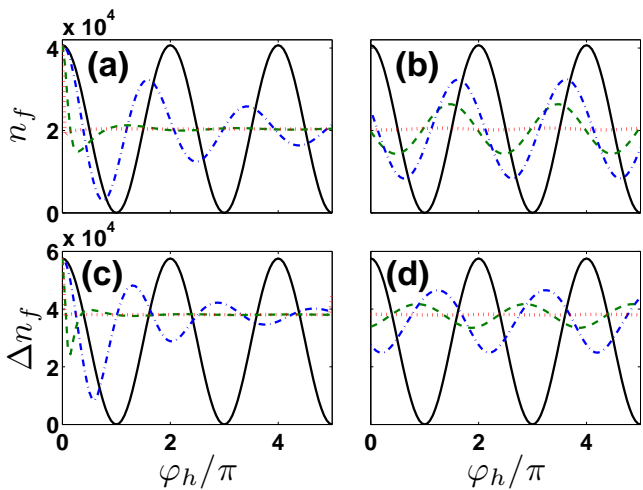}
\caption{(Color online) Final number of atoms $n_f$ (a,b) and its variance $\Delta n_f$ (c,d) as a function of $\varphi_h=\delta t_h$ in an SU(1,1) interferometer with $n=2k(\cosh\theta_p-1)=100$. Time domain  fringes (fixed $\delta$ and varying $t_h$) are shown in (a,c) with $un/\delta=0$ (solid), 0.1 (dash-dotted), 1 (dashed), and 10 (dotted). Frequency domain fringes  (fixed $t_h$ and varying $\delta$) are plotted in (b,d) with $unt_h=0$ (solid), 0.5 (dash-dotted), 1 (dashed), and 10 (dotted).}
\label{ampd_fig3}
\end{figure}
%%%%%%%%%%%%%%%%%%%%%%

%%%%%%%%%%%%%%%%% FRINGES WITH INTERACTIONS  %%%%%%%%%%%%%%%%%
To demonstrate the effect of interactions on the SU(1,1) interferometer, we find the final atom number $n_f(\varphi_h)$ with phase-disffusion present during the hold time,
\begin{equation}
n_f=2k\left\{1+\frac{\cos\Phi_h}{\left[1+\sin^2(ut_h)\sinh^2\theta_p\right]^{k+1/2}}\right\}\sinh^2\theta_p~,
\label{nfpd}
\end{equation}
where $\Phi_h=\varphi_h+(2k+1)\arctan[\cosh\theta_p\tan(ut_h)]$.  An exact form is also found for $\Delta n_f$. The Ramsey-like fringes are thus shifted due to the collisional shift in the atomic energy, and attenuated due to the loss of atom-molecule coherence (Fig.~\ref{ampd_fig3}). They vanish on a $\tau_{pd}$ timescale, approaching the fixed value $n_f=2k\sinh^2\theta_p$ (which corresponds to the state depicted by a dotted ellipse in Fig. ~\ref{ampd_fig1}(d)). It is also evident from Eq. (\ref{pdsu11}) and Eq. (\ref{nfpd}) that coherence revives on a very long $\tau_r=\pi/u$ timescale, similarly to the SU(2) case \cite{PhaseDiffusion,PDexperiments}.
%%%%%%%%%%%%%%%%%%%%%%%%%%%%%%%%%%%%%%%%%%%%%%%%%%%

%%%%%%%%%%%%%%%%%  CONCLUSIONS  %%%%%%%%%%%%%%%%%%%%%%%%%
To conclude, the dissociation of molecular BECs holds great potential for the construction of Heisenberg limited SU(1,1) interferometers, due to the inherent phase-squeezing of the TACS. However, phase-squeezing comes at the price of a super-Poissonian $\Delta n\sim n$ number distribution, making the TACS very sensitive to collisional phase-diffusion. The same observation holds true for the SU(2) phase-squeezed states produced by rotation of number-squeezed inputs, in proposals for sub-shot-noise Mach-Zendher atom interferometry \cite{SU11Interferometer,MachZendherSqueezed}. Controlling this dephasing process will pose a major challenge to the implementation of precise atom interferometers, as well as to the realization of coherent superchemistry \cite{SuperChemistry,StimulatedDissociation}.
%%%%%%%%%%%%%%%%%%%%%%%%%%%%%%%%%%%%%%%%%%%%%%%%%%%

This work was supported by the Israel Science Foundation (Grant 582/07).

%%%%%%%%%%%%%%%%%%%%%%%%%%%%%%%%%%%%%%%%%%%%%%%%%%%%%%%%%%%%%%%%%%%%%%%%%%%%%%
%%%%%%%%%%%%%%%%%%%%%%%%%%%%%%%%%%%%%%%%%%%%%%%%%%%%%%%%%%%%%%%%%%%%%%%%%%%%%%

%%%%%%%%%%%%%%%%%%%%%%%%%%%%%%%%%%%%%%%%%%%%%%%%%%%%%%%%%%%%%%%%%%%%%%%%%%%%%%
%%%%%%%%%%%%%%%%%%%%%%%%%%%%%%%%%%%%%%%%%%%%%%%%%%%%%%%%%%%%%%%%%%%%%%%%%%%%%%

\end{document}